\documentclass[conference]{IEEEtran}
\IEEEoverridecommandlockouts
\usepackage{cite}
\usepackage{amsmath,amssymb,amsfonts}
\usepackage{algorithmic}
\usepackage{graphicx}
\usepackage{textcomp}
\usepackage{xcolor}
\def\BibTeX{{\rm B\kern-.05em{\sc i\kern-.025em b}\kern-.08em
    T\kern-.1667em\lower.7ex\hbox{E}\kern-.125emX}}
\begin{document}

\title{Memory Wall is not gone: A Critical Outlook on Memory Architecture in Digital Neuromorphic Computing}

\author{Amirreza Yousefzadeh, Sameed Sohail, Ana Lucia Varbanescu\\
University of Twente, Enschede, NL\\
}

\maketitle
\begingroup
\renewcommand\thefootnote{}\footnotetext{%
\hspace{-0.5em}%
© 2025 IEEE. Personal use of this material is permitted. 
Permission from IEEE must be obtained for all other uses, in any current or future media, including reprinting/republishing this material for advertising or promotional purposes, creating new collective works, for resale or redistribution to servers or lists, or reuse of any copyrighted component of this work in other works. DOI: 10.1109/ISVLSI65124.2025.11130262}%
\addtocounter{footnote}{1}%
\endgroup

\begin{abstract}
The rapid advancement of neuromorphic technology aims to address the memory wall challenge inherent in conventional von Neumann architectures. This paper critically examines current digital neuromorphic processors and their strategies to mitigate this bottleneck. While designed to bring computation closer to memory through distributed architectures, our findings indicate that on-chip memory systems, including SRAM and emerging technologies like STT-MRAM, have become significant consumers of area and energy, leading to a new memory wall. Through an analysis of energy and area efficiency in various memory technologies, we argue that without a re-evaluation of memory organization, digital neuromorphic processors may struggle to compete effectively in edge and embedded applications. We conclude with potential pathways for future research to overcome the limitations of on-chip memory in neuromorphic systems.
\end{abstract}


\section{Introduction}\label{sec:intro}

For the past decade, the performance of von Neumann machines has been limited not by their arithmetic capabilities, but by the cost of moving data between cores and main memory. In modern AI accelerators, the energy required to fetch a single 8-bit weight from off-core SRAM often exceeds the energy used in a multiply-accumulate (MAC) operation by more than an order of magnitude\cite{tang2023open}. Additionally, latency issues are heavily influenced by the memory hierarchy. This phenomenon, commonly referred to as the "memory wall," has become the primary bottleneck for both throughput and overall system energy efficiency\cite{wulf1995hitting}.

\begin{figure}
    \centering
    \includegraphics[width=0.6\linewidth]{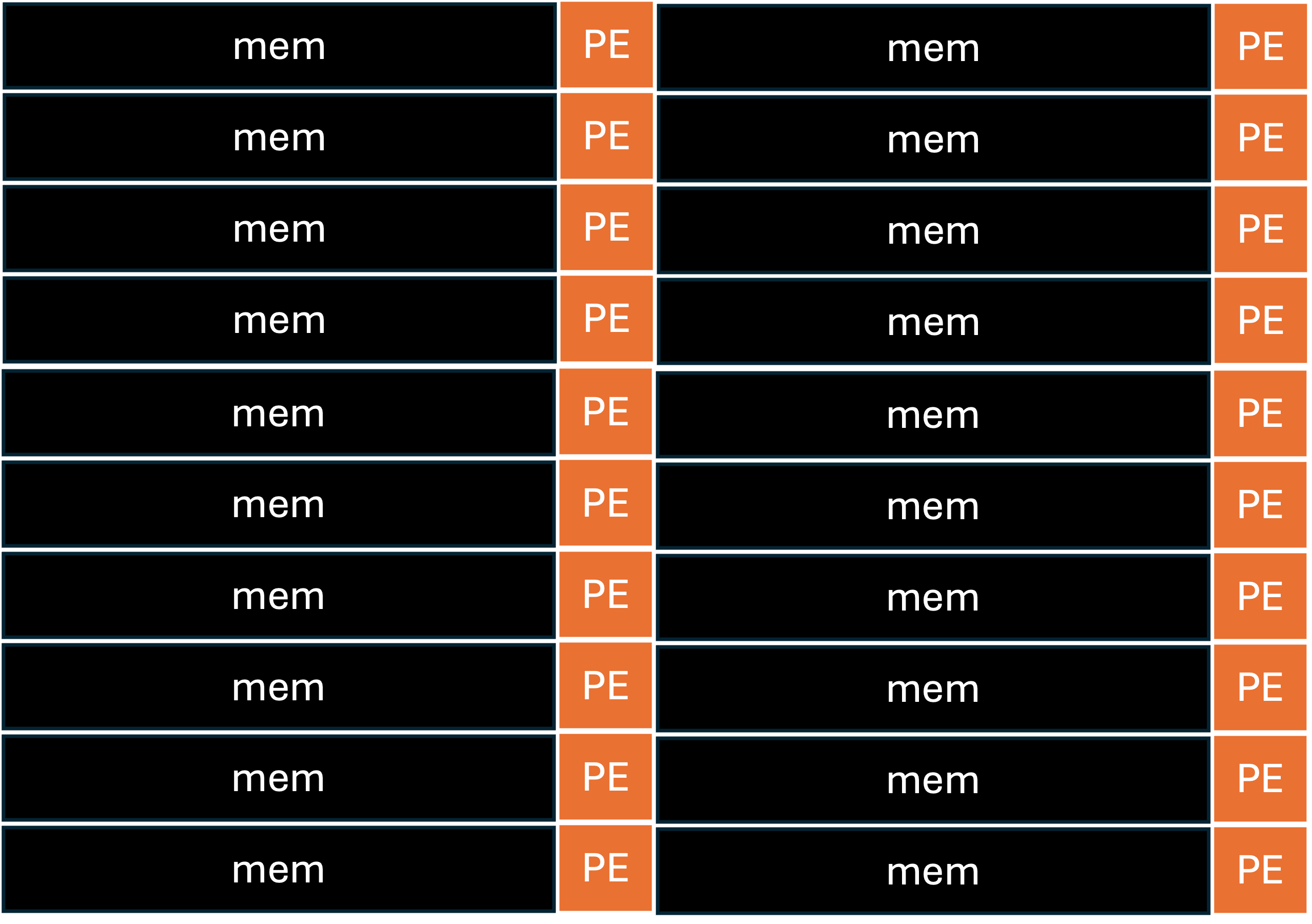}
    \caption{Fully distributed memory (mem) and processing elements (PE) in neuromorphic processors aim to eliminate the memory wall.}
    \label{fig:distributed_memory}
\end{figure}

Neuromorphic architectures draw direct inspiration from biological systems, where each biological neuron stores its parameters locally. Applying this concept to silicon suggests the development of a tiled, fully distributed architecture. In this design, small memory slices are positioned closely alongside lightweight processing elements (PEs), as illustrated in Fig.~\ref{fig:distributed_memory}. Theoretically, this co-location could eliminate the classical memory wall by reducing the need for costly off-chip and off-core data transfers.

\begin{figure}
    \centering
    \includegraphics[width=1\linewidth]{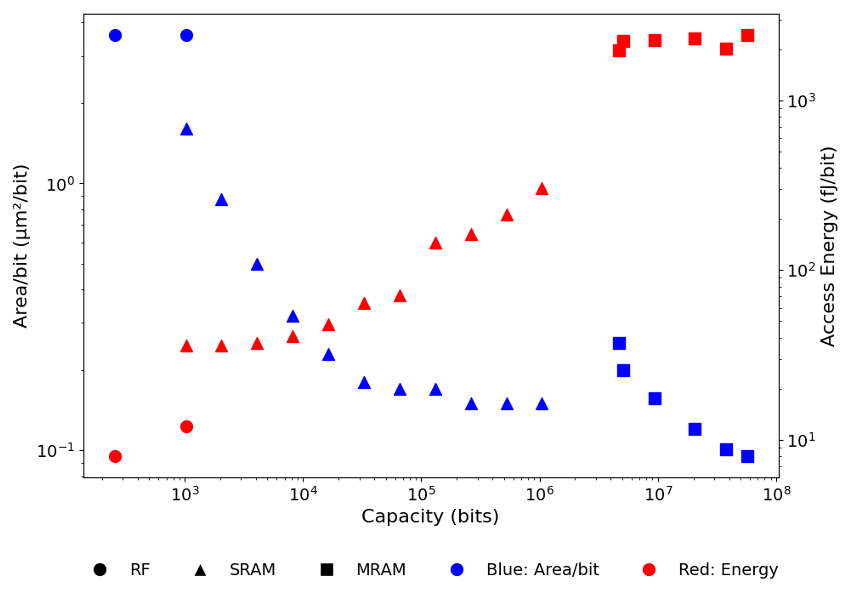}
    \caption{The trade-off between area efficiency and energy consumption in three different integrated (on-chip) memory technologies at the FDX-22nm node, including Register-Files (RF) created using standard digital cells, SRAM blocks, and MRAM blocks designed with commercial memory compilers. For the MRAM, the energy consumption mentioned refers solely to read access.}
    \label{fig:memory_area_energy}
\end{figure}

As shown in Fig.~\ref{fig:memory_area_energy}, distributed memories tend to consume less power but occupy significantly more silicon area. While smaller, energy-efficient memories come with significant density penalties due to peripheral overhead, dense SRAM and MRAM macros can consume 10 to 100 times more energy per access. Additionally, low area efficiency can also indirectly impact power efficiency, as the increased distance between transistors raises energy consumption. In state-of-the-art digital neuromorphic chips, on-chip memories are the primary consumers of both silicon area and energy \cite{tang2023seneca}. As a result, modern neuromorphic processors have replaced the traditional memory wall with a new memory wall, driven by the area and energy limitations of on-chip storage.

In this paper, we argue that digital neuromorphic processors will not be competitive for the edge and embedded workloads they are designed for unless their memory organization is rethought. Currently, computation is no longer the bottleneck; instead, memory has become the limiting factor.

\section{Energy--Area Trade-offs in Memory Architectures}
\label{sec:memtech}

\begin{figure}
    \centering
    \includegraphics[width=0.8\linewidth]{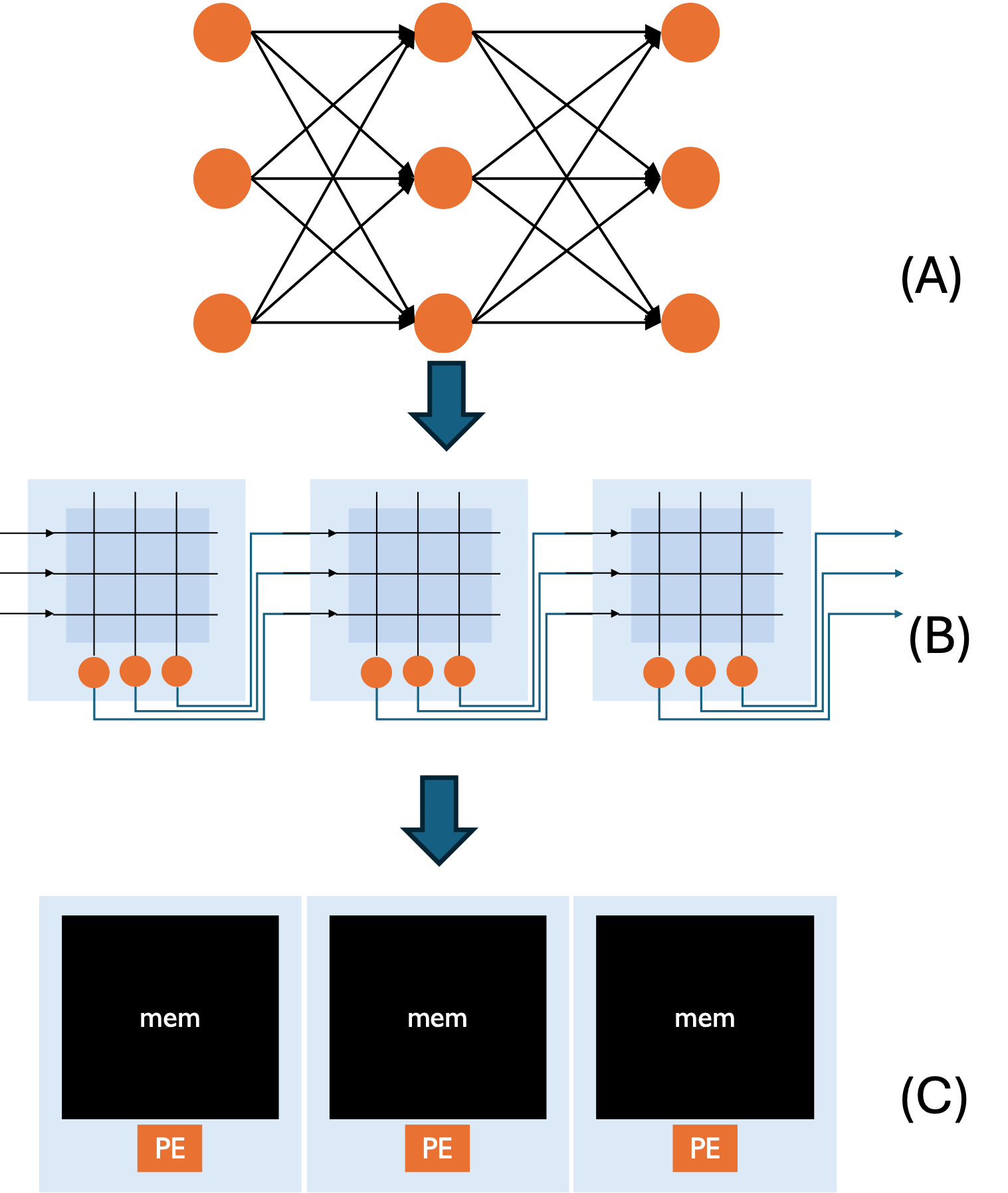}
    \caption{Spatial mapping of the layers of a neural network onto tiled memory-processing element (PE) cores.}
    \label{fig:neuromorphic_processor}
\end{figure}

The biological brain stores and processes memory in an exceptionally distributed manner: every synapse is physically connected to its pre- and post-synaptic neurons. This means that moving a charge just a few hundred nanometers allows for both storage and computation to occur simultaneously. Achieving this level of granularity in digital technology would necessitate trillions of individually addressable (programmable) single-bit memories. Instead, existing digital neuromorphic processors group tens of thousands of synapses into regular SRAM or crossbar macros, which can be replicated as tileable cores (see Fig.~\ref{fig:neuromorphic_processor}). This grouping of bits helps optimize memory peripheral overheads, significantly improving area efficiency and facilitating large-scale integration.

Grouping synapses into memory blocks allows for time-multiplexing of neural computing logic, further enhancing area efficiency (Fig. \ref{fig:neuromorphic_processor}.C). Digital neuromorphic chips utilize GHz-rate CMOS technology to perform time-multiplexed computation. This means that a single physical Processing Element (PE) can emulate hundreds of logical neurons each millisecond, significantly reducing the logic area by more than an order of magnitude. However, memories cannot be time-multiplexed since each logical neuron requires its parameters and state to be retained at all times. This leads to a significant imbalance: in recent chips, memory occupies more than 80\% of the die area and accounts for the majority of both dynamic and static power consumption.

As the capacity of memory blocks increases, the energy required for access grows approximately logarithmically with the length of the word line. Additionally, the physical distance between communicating tiles increases both the network energy consumption and latency. This creates a challenging trade-off for designers between area and energy efficiency. Fig.~\ref{fig:memory_area_energy} plots measured points from a 22\,nm FDX process for three mainstream on-chip memories:

\begin{itemize}
  \item Small Register Files built from standard-cell flops: offer the lowest access energies ($<5$\,fJ/bit) because bit-lines and word-lines are only a few microns long, but they occupy significant silicon area per bit, yielding the \emph{worst} density ($>2\,$µm\(^2\)/bit).
  \item Compiled SRAM macros from 1Kb to 1Mb:  density improves roughly inversely with capacity because peripheral overhead is amortised; however, the longer wires increase \(C\!V^2\) switching loss so energy/bit \textit{rises} from $\sim5$fJ to $\sim80$fJ.
  \item Non-Volatile MRAM macros from 1Mb to 100Mb: offer an attractive sub-0.1µm\(^2\)/bit at capacities $>80$Mbit, but each read already costs thousands of fJ; writes are 20 to 30 times higher and therefore excluded from the plot.
\end{itemize}

Reducing access energy by dividing memories into smaller sizes can significantly increase the area required per bit. Additionally, this approach may even lead to an indirect rise in power consumption due to the following reasons:
\begin{itemize}
    \item Overloading core-to-core communication, resulting from the implementation of a higher number of cores with smaller memories.
    \item Increasing leakage, as static power consumption tends to scale roughly with the silicon area.
\end{itemize}

Additionally, larger die sizes are significantly more expensive due to lower yields, influencing the energy-optimal point for memory capacity within neuromorphic designs to be less appealing for cost-sensitive applications.


Efforts to enhance energy efficiency through memory fragmentation merely shift the problem to other areas, leading to increased power consumption due to leakage, communication between tiles, and less efficient mapping. Conversely, while packing larger macros may enhance area efficiency, it results in higher energy consumption per access. Neither of these strategies addresses the fundamental issue.

\section{Mapping Inefficiency in Neuromorphic Processors}\label{sec:sotachips}

Table~\ref{tab:sota} presents an overview of several prominent digital neuromorphic platforms. While each design relies on low-leakage SRAM, the percentage of bits used to store \emph{useful} model parameters typically does not exceed 30\%. In highly distributed architectures, this proportion can drop to below 1\%.

\begin{table*}
  \centering
  \caption{Memory utilisation in state-of-the-art neuromorphic chips. Mapping efficiency is defined as the ratio of parameter bits to the aggregate on-chip memory capacity allocated to the network.}
  \label{tab:sota}
  \begin{tabular}{|l|c|c|c|c|c|c|}
  \hline
  \textbf{Architecture} & \textbf{Synapse} & \textbf{Neuron} & \textbf{Benchmark} &
  \textbf{\#Params} & \textbf{On-chip} & \textbf{Mapping} \\
  & \textbf{Memory} & \textbf{Memory} & \textbf{Network} &
  \textbf{(bits)} & \textbf{Mem.\ (bits)} & \textbf{Efficiency} \\
  \hline
  IBM TrueNorth \cite{doi:10.1073/pnas.1604850113}         & SRAM & SRAM & CIFAR-10   & 2M (2.3Mb)  & 440\,M  & 0.5\,\% \\
  Intel Loihi  \cite{rueckauer2022nxtf}          & SRAM & SRAM & MobileNet & 4M (36Mb)   & 4.0\,G  & 0.9\,\% \\
  GrAI VIP \cite{zhu2022arts}              & SRAM & SRAM & MobileNet & 3.07M (24.56Mb) & 89.8\,M & 27.3\,\% \\
  SPECK \cite{richter2023speck}                  & SRAM & SRAM & N-MNIST   & 9.3K (74.4Kb) & 250\,k  & 29.8\,\% \\
  \hline
  \end{tabular}
\end{table*}


The ideal ratio of model parameters to mapped memory bits should be close to 100\%. However, in practice, this ratio rarely exceeds 30\%. The reasons for this include: (i) the fact that cores come in discrete sizes that seldom align with the layer dimensions of a specific network, and (ii) High-resolution neuron states often consume as many bits as the weights themselves. 

Neuromorphic processors consist of thousands of small, fixed-size SRAM slices. When a layer does not fully utilize a core, the unused space results in dark silicon, which is a wasted area. Additionally, when a layer extends across multiple cores, it spreads the communicating neurons, leading to an increase in the number of router hops and greater power consumption in the spike network. This explains the low mapping efficiency of the first two rows in Table~\ref{tab:sota}: deploying a MobileNet with 36 Mb of parameters on Loihi requires 4Gb of on-chip SRAM (16 chips), leading to a mapping efficiency of less than 1\% \cite{rueckauer2022nxtf}. The higher mapping efficiencies of SPECK\cite{richter2023speck} stem from SPECK's dedicated and less distributed architecture, which reduces memory fragmentation.

Unlike conventional DNN accelerators, neuromorphic processors must store high-precision neuron states (e.g., membrane potentials) continuously across all network layers. While DNN accelerators typically use lower-precision and transient partial sums to reduce memory demands, neuromorphic chips employ high-resolution representations for neuron states to avoid overflow during repeated accumulations of synaptic weights. Such persistent, high-precision storage dramatically inflates memory usage. The higher mapping efficiencies of GrAI VIP\cite{zhu2022arts} derive from GrAI-VIP's hybrid design, which allows for the implementation of stateless neurons.

The data presented in Table~\ref{tab:sota} indicate that current neuromorphic chips do not just encounter the limitations of on-chip memory; they crash into it, resulting in a waste of 70–99\% of the available bits. In reality, the situation is even more severe, as the published benchmarks for neural networks on these platforms are often customized extensively to accommodate the hardware constraints.

\section{Discussion and Solution Space}
\label{sec:solutions}

Addressing the on-chip memory wall in digital neuromorphic processors demands a comprehensive approach that integrates innovations across memory technology, architecture, and algorithm design. Below, we outline several promising directions and practical solutions.

\subsection{Algorithm: Hybrid neural networks}

Conventional spiking neural networks (SNNs) assign a persistent, high-precision membrane state to every neuron in each layer. This all-stateful approach maximizes temporal expressiveness but increases memory usage and static power consumption. A more memory-efficient alternative is the hybrid neural network, which combines stateful spiking layers only in areas where temporal dynamics are beneficial, while using stateless feedforward layers in other parts of the network pipeline \cite{arjmand2024trip, Shenqi_CEED}.

\subsection{Software: Smart Scheduling}
Software-level optimizations can greatly reduce memory bandwidth requirements. Techniques such as spike grouping\cite{xu2024optimizing} effectively decrease the volume of data that needs to be transferred and stored. Spike grouping takes advantage of both temporal and spatial locality by batching events, which reduces the number of memory transactions. 

\subsection{Architecture: Heterogeneous and Hierarchical Memories}
An effective approach to address memory inefficiencies is to integrate heterogeneous memory technologies organized hierarchically. This would involve using very small, low-energy register files (RF) for frequently accessed neuron states (“hot” data), medium-sized SRAM blocks optimized for weights, and denser non-volatile memories (NVM) such as MRAM or RRAM for infrequently updated parameters (“cold” data) \cite{simon_farshad}. Such a structure offers a flexible balance, optimizing both energy and area efficiency. 

For event-based systems with numerous parallel processing elements, introducing a heterogeneous and possibly hierarchical memory system presents a challenge. A few neuromorphic processors, such as Seneca \cite{xu2024optimizing} and SpiNNaker \cite{hoppner2021spinnaker, furber2014spinnaker}, allow the use of off-chip DDR memory to facilitate large-scale neural network deployment. However, opting for off-chip memory significantly increases power consumption and reduces the overall performance of the system \cite{yousefzadeh2018performance, nembhani2024sensim}. Unlike heterogeneous computing, which utilizes different types of processing elements, research on heterogeneous memory systems in neuromorphic computing remains quite limited.

\subsection{Technology: In-Memory Compute}

Non-volatile memory (NVM) technologies, such as MRAM and RRAM, provide a promising capability known as \emph{in-memory compute}. This allows synaptic multiplication operations to occur directly within the memory array, which significantly reduces the number of memory accesses required. Typically implemented in crossbar architectures, these operations enable the weights stored in memory to directly interact with incoming input signals (or spikes). However, the subsequent addition of these partial sums to neuron states still takes place in peripheral circuitry rather than in the memory cells themselves \cite{gebregiorgis2023tutorial}. This limitation restricts the overall energy savings that can be achieved through this approach.

Moreover, in-memory compute architectures impose strict constraints on neural network mapping, as crossbar dimensions are typically fixed and inflexible \cite{bhattacharjee2020crossbar}. This rigid structure frequently leads to suboptimal memory utilization, reducing mapping efficiency and necessitating replication or padding of neural network weights. Consequently, the area overhead increases significantly, potentially offsetting energy advantages and limiting overall practicality for general-purpose neuromorphic computing \cite{khaddam2022hermes}.

Another approach involves integrating processing capabilities within DRAM dies, known as processing-in-memory (PIM) \cite{sudarshan2022critical}. This method embeds simple processing units within or adjacent to DRAM arrays, enabling computations to occur closer to where data resides. While PIM can alleviate some data movement bottlenecks, accessing large DRAM blocks, even with integrated processing, still incurs significant energy costs compared to on-chip SRAM. This method is only effective in architectures that utilize a hierarchical memory system.

\subsection{Technology: 3-D Integration with Monolithically-Integrated NVM Layers}
Monolithic 3-D integration is a promising technology for developing distributed memories on a chip. In this method, one or more layers of non-volatile memory (NVM)—such as STT-MRAM, ReRAM, or PCM—are fabricated sequentially in the back-end-of-line (BEOL) metal tiers above a completed CMOS logic wafer \cite{rocha2024multidie, ahmadi2022hybrid}. 

Placing non-volatile memory (NVM) tiles in the upper metal layers creates more space on the front-end silicon for logic circuits and other types of memory, such as SRAM. Vertical vias minimize the distance that signals must travel to just a few microns, which reduces both the dynamic energy consumed by interconnects and latency. Monolithic 3-D NVM offers a pathway to integrate memory and computation into a cohesive, brain-like vertically integrated structure. However, it does not eliminate the need for a heterogeneous memory architecture, as it still suffers from most of the disadvantages of non-volatile memories.

\bibliographystyle{IEEEtran}
\bibliography{references}

\end{document}